\documentclass[preprint, amssymb,  aps, showpacs,preprintnumbers,superscriptaddress,
amsmath,showkeys,floatfix]{revtex4}

\usepackage{epstopdf}
\usepackage{graphics}
\usepackage{graphicx}
\usepackage{dcolumn}
\usepackage{bm}
\usepackage{longtable}
\usepackage{epsfig}
\usepackage{times}
\usepackage{url}
\usepackage{color}

\begin{document}
\title{Nuclear recollisions in laser-assisted $\alpha$ decay}

\author{H\'ector Mauricio \surname{Casta\~neda Cort\'es}}
\affiliation{Max-Planck-Institut f\"ur Kernphysik, Saupfercheckweg 1, D-69117 Heidelberg, Germany}

\author{Carsten \surname{M\"uller}} 
\affiliation{Max-Planck-Institut f\"ur Kernphysik, Saupfercheckweg 1, D-69117 Heidelberg, Germany}
\affiliation{Institut f\"ur Theoretische Physik I, Heinrich-Heine-Universit\"at D\"usseldorf, Universit\"atsstr. 1, D-40225 D\"usseldorf, Germany}

\author{Christoph~H. \surname{Keitel}}
\affiliation{Max-Planck-Institut f\"ur Kernphysik, Saupfercheckweg 1, D-69117 Heidelberg, Germany}

\author{Adriana \surname{P\'alffy}}
\email{Palffy@mpi-hd.mpg.de}
\affiliation{Max-Planck-Institut f\"ur Kernphysik, Saupfercheckweg 1, D-69117 Heidelberg, Germany}

\date{\today}
\begin{abstract}

Laser-induced nuclear recollisions following $\alpha$ decay in the presence of an intense laser field are investigated theoretically.
We show that while an intense optical laser does not influence notably the tunneling rate in $\alpha$ decay, it can completely change the
$\alpha$ particle spectrum. For intensities of $10^{22}-10^{23}$ W/cm$^{2}$, the field is strong enough to induce recollisions between the
emitted $\alpha$ particle and the daughter nucleus. The energy gained by the $\alpha$ particle in the field can reach 20 MeV and suffice
to trigger several types of nuclear reactions on a femtosecond time scale. Similar conclusions can be drawn about laser-induced recollisions after proton emission. Prospects for the experimental realization of laser-induced nuclear recollisions are discussed.

\end{abstract}
\maketitle

Laser-driven recollisions have come to play a crucial part in atomic strong-field physics, due to their role  in non-sequential double ionization \cite{LHuillier83,Fittinghof92} and high-harmonic generation (HHG) \cite{McPherson87}. In turn, HHG has opened the field of attoscience, and  the emergence of extremely short pulses 
has rendered possible the control of electron dynamics in atoms and ions \cite{Scrinzi2006,Kienberger2007,Nisoli2009,Krausz2009,Brif2010}. So far, recollision studies addressed  laser-driven electrons \cite{Dieter2006,reviewCHK2012} or muons \cite{Atif2007} returning to the emitting ion, i.e., rescattering of leptons on an attractive Coulomb potential.
The high recollision energies reached by rescattering muons could allow  to probe the nuclear structure due to the smaller Bohr radius of the bound muon \cite{Atif2007}. 
Closer to nuclear physics applications, recollisions and muon-catalyzed fusion in the short-lived muonic $\mathrm{D}^+_2$ molecular ion in the presence of a superintense laser field  were investigated \cite{Corkum2004}.

In this work we investigate for the first time laser-driven recollisions of $\alpha$ particles following $\alpha$ decay. Our study  complements to the fields of direct and indirect interaction of coherent light with nuclei which comprise for instance coherent driving of nuclear transitions \cite{olga,nqo,dipfor,nstirap,gdrHans}, electron bridge mechanisms in laser-assisted internal conversion \cite{elbridge} or the laser-assisted $\beta$ decay \cite{beta} and  nuclear photoeffect \cite{photoef}.
A charged heavy particle bound by the strong force, the $\alpha$ particle tunnels through the nuclear and Coulomb barrier of the nucleus, as was first described in 1928 by Gamow \cite{Gamow28} and independently by Condon and Gurney \cite{ConGur28}. Under the action of a strong laser field, the emitted $\alpha$ particle may change its trajectory and recollide with the daughter nucleus at energies sufficient to produce nuclear reactions and on time scales that are so far not available in experiments. Such fast recollisions can even allow probing short-lived excited nuclear states reached via $\alpha$ decay. Thus, laser-driven nuclear recollisions in $\alpha$ decay are not only a different physical system for the study of strong-field effects, involving  a repulsive potential, but also open the possibility to investigate a new energy regime which hosts the interplay between the electromagnetic and the strong force.   We show here that such recollisions are rare but detectable already at presently available laser intensities of $10^{22}-10^{23}$ W/cm$^{2}$.

Due to the heavy mass of the $\alpha$ particle compared to that of the electron, $\alpha$ decay is an excellent example of a  non-relativistic process in the semiclassical parameter regime. In order to study how strong-laser pulses affect  $\alpha$ decay, two of us have developed a  formalism that can treat the  laser-assisted tunneling of quasi-stationary states \cite{lat2011}. Starting from the so-called Strong-Field Approximation (SFA) and its formulation in terms of trajectories in imaginary time, our method describes both qualitatively and quantitatively the tunneling of quasistationary states in laser fields in the semiclassical parameter regime. While in Ref.~\cite{lat2011} this method was applied for the test case of a short-range potential, in order to describe laser-assisted $\alpha$ decay we need to  generalize this result for the long-range Coulomb barrier \cite{HectorDiss2011}. The approach we have adopted is one dimensional following  successful models that have proven their predictive power in non-relativistic laser-atom interactions \cite{Eberly}.

\begin{figure}[b]
\vspace{-0.4cm}
  \includegraphics[width=10cm]{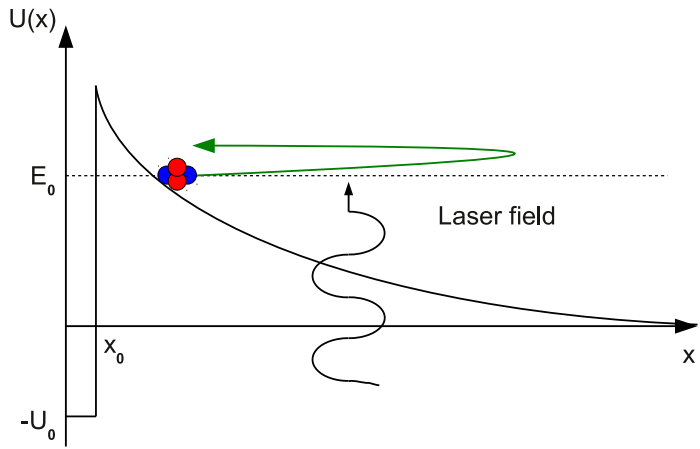}
  \caption{\label{fig1}
 Qualitative illustration of the 1D nuclear barrier as described in Ref.~\cite{Buck}. Nuclear forces are responsible for the well region $\mathrm{x}\ll \mathrm{x}_0$, while the barrier is determined solely by the Coulomb interaction. 
  }
\end{figure}

Our  theoretical formalism considers the field-assisted tunneling of the preformed $\alpha$ particle through the Coulomb barrier of the nucleus, following  the framework of the precluster model \cite{Buck}. The preformed $\alpha$ cluster is initially confined in a potential well with depth $-\mathrm{U}_0$, which is taken as the mean field nuclear potential that the nucleons of the parent nucleus experience. The nuclear interaction is short-ranged, as shown in Fig.~\ref{fig1}, so the potential well has a finite length x$_0$. For distances larger than x$_0$, the  interaction is dictated by the Coulomb force acting between the protons of the daughter nucleus and the $\alpha$ particle. The interaction of the $\alpha$ particle with the atomic electrons is neglected.
The parameters $\mathrm{U}_0$ and x$_0$ are calculated following the method described in Ref.~\cite{Buck,Royer}. According to the imaginary time method (ITM) \cite{popov-itm}, a trajectory x$(t)$ can be found along which the particle starts its motion at the complex time instant $t=t_s$ inside the well, x$(t_s)=0$, having the energy $E_0$ and arrives at the exit of the barrier when $t=t_0$. 
The trajectory satisfies the Newton equation $m_r\ddot{\mathrm{x}}=-{\partial U}/{\partial \mathrm{x}}+\dot p_F(t)$,  
where $m_r$ is the reduced mass of the nuclear system composed of $\alpha$ particle and daughter nucleus, $p_F(t)$ is the laser-induced momentum  $p_F(t)=Z_{\alpha}eA(t)/c$,  $Z_{\alpha}$ is
the charge number of the $\alpha$ particle, $e$ is the charge of the proton, $A(t)$ is  the one-dimensional 
field vector potential and $U(\mathrm{x})$ is the total nuclear potential. 
The exit point is separated from the well by the classically forbidden region, so that the solution of the Newton equation satisfying the assigned initial conditions only exists in complex time, $t=t_0+i\tau$. After exiting the barrier, the $\alpha$ particle moves under the action of both  long-range Coulomb potential and  electric field of the laser up to a detector placed far away from the emitting nucleus. The tunneling rate can be written with the help of the classical action $W$  along the complex trajectories, starting from the modified SFA transition amplitude \cite{lat2011}.

\begin{figure}[!ht]
\includegraphics[width=11cm]{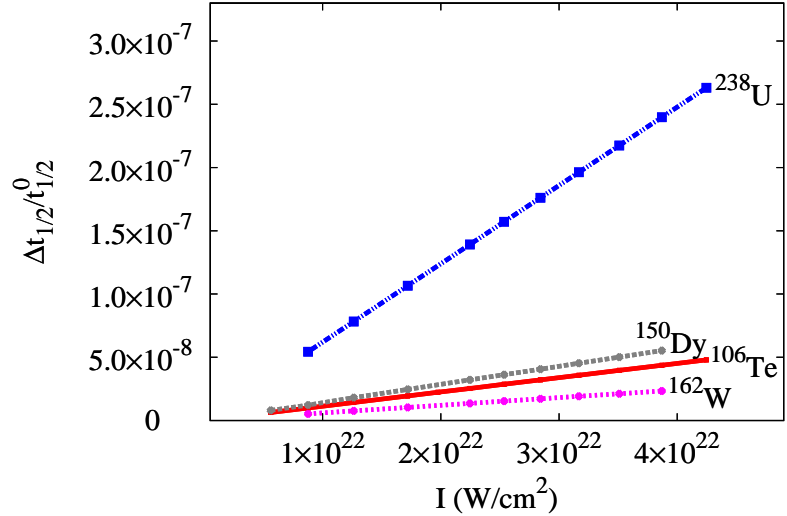}
\caption{\label{fig4:laserlifetimesintensities} (Color online) Relative modification of the laser-assisted half-lives $\Delta t_{1/2}/t_{1/2}^0=\left(t_{1/2}^0-t_{1/2}(I)\right)/t_{1/2}^0$ as a function of the laser intensity $I$.  The considered laser frequency is 800 nm (Ti:Sa laser) corresponding to the  photon energy $\hbar\omega=1.55$ eV.}
\end{figure}

For an estimate of the laser effect on the $\alpha$ decay rate, we have first considered the idealized case of a strong monochromatic field.
The transition amplitude as a function of the $\alpha$ particle momentum $p$ in this case is given by 
\begin{eqnarray}
\label{transitionamplitudemonochromatic_alpha}
M(p)&=& \hbar\omega\underset{l=-\infty}{\overset{l}{\sum}}\delta\Bigl(\frac{p^2}{2m_\text{r}}+\frac{Z_\alpha^2e^2\mathcal{E}_0^2}{4m_\text{r}\omega^2}-E_0- l \hbar \omega\Bigr)
\nonumber \\
&\times&
\sum_{\eta}\frac{\mathcal {P}_{0}\exp\left[iW(p,t_{s\eta})\right]}{\sqrt{dp/dt_{0\eta}+i\beta}}\, ,
\end{eqnarray}
where   $\mathcal{P}_{0}^2$ the pre-exponential factor of the field-free decay as given by the Wentzel-Kramers-Brillouin (WKB) formula  \cite{lat2011} and  $\beta$  the regularization constant needed to avoid  the divergencies at the classical cut-offs where two or more trajectories meet \cite{Carla,Sergey}. Furthermore, $\mathcal{E}_0$ is the electric field strength  and $\omega$ the laser frequency. The total action $W$ is evaluated at the saddle-point initial times  $t_{s\eta}$  and the final result is obtained by summing over all saddle points $\eta$ \cite{popov-itm}. This corresponds to summing over the complex trajectories that the particle can follow through the tunneling barrier. In the case of no recollisions, there are only two such trajectories for each final energy  of the $\alpha$ particle arriving at the detector.
The spectrum of the $\alpha$ particles, i.e., the differential decay rate $dR/dp$, is  given by the modulus square of the transition amplitude $M(p)$ in Eq.~\eqref{transitionamplitudemonochromatic_alpha} evaluated on the long time of observation $T$, 
\begin{eqnarray}
dR&=&\frac{|M(p)|^2dp}{T}=\sum_{j}\frac{\delta(p-p_j)}{2\pi p_j}\mathcal{P}_0^2\hbar^2\omega^2
\nonumber \\
&\times&
\left\vert\sum_{\eta}\frac{\exp\left(iW(p,t_{s\eta})\right)}
{\sqrt{dp/dt_{0\eta}+i\beta}}\right\vert^2dp\, ,
\label{therate_alpha}
\end{eqnarray}
where $p_j=\sqrt{2m_\text{r}\left(E_0-Z_\alpha^2e^2\mathcal{E}_0^2/(4m_\text{r}\omega^2)+j\hbar \omega\right)}$ are the momenta corresponding to the above-threshold ionization (ATI)-like peaks. We find that the total decay rate given by the integral over all final momenta coincides with a very good accuracy with the  WKB field-free decay rate obtained in Ref.~\cite{Buck} using the same barrier parameters.  Fig.~\ref{fig4:laserlifetimesintensities}  presents the laser-assisted half-lives in the case of $^{106}$Te, $^{150}$Dy $^{162}$W and $^{238}$U as a function of several laser intensities. Our results show that the laser effect is to accelerate the decay. However, the relative modification of the natural half-lives $t_{1/2}^0$ is extremely small, on the order of $ 10^{-7}-10^{-8}$, showing a linear dependence on the field intensity $I$. This has been discussed in Ref.~\cite{lat2011} for the case of tunneling through a short-range potential 
 and can be traced back to the field-induced modification in the action $W$ which is in the first (relevant) order given by  $\mathcal{E}_0 b^2/p_0$, with $b$ the barrier thickness and $p_0$ the initial momentum. For the $\alpha$ decay parameters this factor is small enough to reduce the exponential dependence in Eq.~(\ref{transitionamplitudemonochromatic_alpha}) into a linear one as seen in  Fig.~\ref{fig4:laserlifetimesintensities}. The slope is approx. given by $\hbar^2b^4\exp(-2bp_0/\hbar)/(4p_0^2)$ and is steepest for the case of  $^{238}$U, due to its largest barrier thickness $b$ among the studied $\alpha$-emitters.

%
%

While the effect of the laser field on the tunneling rate itself turns out to be negligible, the spectrum of the emitted $\alpha$ particles is strongly modified by the laser field.
The problem can be safely treated non-relativistically since the value of the relativistic field strength parameter \cite{param,rev} $\xi=Z_{\alpha}e\mathcal{E}_0/(m_{\rm r}\omega c) \simeq 0.05$  ($c$ denotes here the speed of light) is much smaller than one and the magnetically-induced relativistic drift \cite{Palalal} is for the considered cases about one order of magnitude smaller than the spread of the $\alpha$ particle wave packet at the recollision. 
 Instead of monochromatic $\alpha$ particles with energy $E_0$, under the action of the laser $\alpha$ particles reach the detector with energies approx. between $(p_0-p_F)^2/(2m_\text{r})$ (or zero) and $(p_0+p_F)^2/(2m_\text{r})$, with $p_F$ the maximum field-induced momentum $p_F=Z_{\alpha}e\mathcal{E}_0/\omega$. The spectrum is composed of a large number of ATI-like maxima, separated by the momentum corresponding to the energy of the laser photons $\hbar\omega$.
For field strengths larger than $\mathcal{E}_0\simeq \omega\sqrt{2m_\text{r}E_0}/(Z_{\alpha}e)$, recollisions of the emitted $\alpha$ particles with the daughter nucleus may occur. This can be seen by plotting the spatial trajectory of the particle outside the barrier for the first few laser cycles after the particle has emerged in the classically allowed region. Since the initial energies $E_0$ of the $\alpha$ particle are on the order of 4--5 MeV, the recollision threshold for the electric field is approx. $10^{22}$ W/cm$^{2}$. Depending on the field strength and phase, the $\alpha$ particle will hit the Coulomb potential of the daughter nucleus at lower or higher energies than $E_0$.

The  recollision scenario can  be summarized as follows: the $\alpha$ particle tunnels virtually unperturbed through the nuclear barrier. At an arbitrary field phase $\phi=\phi_0$ (we consider here the electric field of the laser  $\mathcal{E}_0\cos (\omega t)$ and $\phi=\omega t$) it  emerges outside the barrier and first rapidly leaves the vicinity of the nucleus due to Coulomb repulsion. Only for certain initial phases $\phi_0$, the energy accumulated in the laser field will suffice  to induce a recollision of the $\alpha$ particle with the nuclear Coulomb barrier. 
 We find for instance that for the laser intensity 7.9$\times 10^{22}$ W/cm$^2$ ($\mathcal{E}_0=1500$ a.u.), recollisions occur only for the $\alpha$ particles emitted with $\phi_0\in[1.4\pi,1.9\pi]$. Similarly to the well-studied case of laser-driven atomic recollisions, the maximum recollision energy is approx. 3.17$U_p$, where $U_p$ is the ponderomotive energy $U_p=Z_{\alpha}^2e^2\mathcal{E}_0^2/(4m_\text{r}\omega^2)$.

Since the laser field does not influence the tunneling 
process, the emitted $\alpha$ particle will emerge outside the barrier isotropically in a $4\pi$ solid angle. 
While our approach is one-dimensional and does not take into account this feature, we may geometrically estimate 
 the fraction of $\alpha$ particles that, under the action of the laser field, 
can recollide with the daughter nucleus. We perform our study on the test case of $^{106}\mathrm{Te}$ $\alpha$ 
decay, chosen for its short half-life $t_{1/2}=7\times 10^{-5}$ s \cite{nds106}. Similarly to the atomic physics 
case, our estimate neglects the possible movement of the daughter nucleus in the field, due to its  heavier mass. 
We find that only $\alpha$ particles emitted in the direction of the field with a narrow angular tolerance can 
recollide with the daughter nucleus. The tolerance solid angle depends on 
the field strength and is for the case of $^{106}\mathrm{Te}$ $10^{-8}$ sr for a field intensity of $10^{22}$ 
W/cm$^{2}$ and $5\cdot 10^{-8}$ sr for a field intensity of $10^{23}$ W/cm$^{2}$. This corresponds to a fraction 
of approx. $10^{-9}$ recollisions out of all emitted $\alpha$ particles. A quantum mechanical estimate based on 
the recolliding alpha particle wave packet size delivers a  somewhat smaller value of $10^{-12}$ for 
the  rescattering fraction  \footnote{Similar numbers are obtained for the probability of HHG in ions \cite{Filho}.}.

In order to calculate the recollision spectrum we need to take into account that in this case there can be more than two saddle-point solutions (or complex trajectories) for a single value of the final energy. Thus, in Eq.~(\ref{therate_alpha}) the sum over the complex trajectories will run now over all regular and recollision trajectories that can lead to a certain final energy of the $\alpha$ particle arriving at the detector. Different regularization parameters $\beta$ need to be taken into account for the regular and the recollision trajectories following the procedure described in Refs.~\cite{Carla,Sergey}. Furthermore, in order to have a more realistic description of the recollision process, we  considered a finite laser pulse shape by introducing a damping factor of the field intensity for the equations of motion of the $\alpha$ particle outside the Coulomb barrier. The parameters for the laser field pulse where chosen such as to agree with the 500 fs pulse duration  (corresponding to approx. 100 laser cycles) of one of the operation modes of the Vulcan laser \cite{vulcan}.

\begin{figure}
\includegraphics[width=10cm]{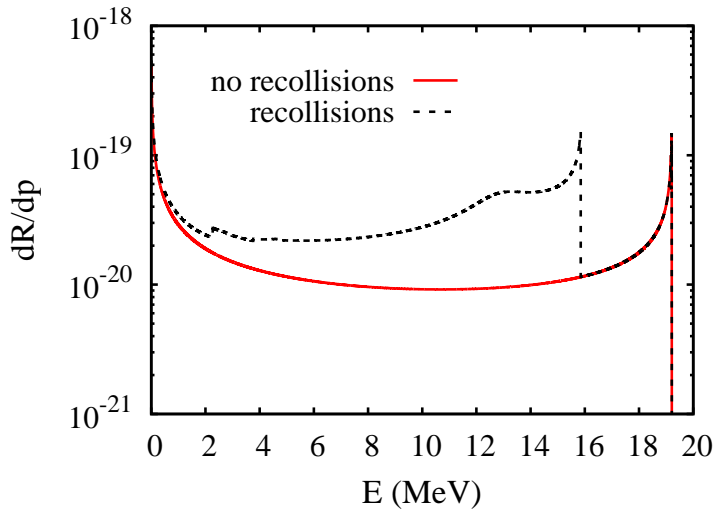}
\caption{(Color online) Energy spectrum of  laser-assisted $\alpha$ decay of $^{106}$Te for $I= 7.9\times 10^{22}$ W/cm$^2$.
The dashed black line depicts the recollision spectrum which is compared with the case of no recollisions for the same laser field strength (solid red line).
  }
\label{recspec2}
\end{figure}

The recollision spectrum for an electric field intensity 7.9$\times 10^{22}$ W/cm$^2$  obtained from our model is presented in Fig.~\ref{recspec2}. For illustration purposes we have discarded   the real part of the action $W$ which would merely introduce very fast oscillations about the spectrum shape.  The recollision trajectories contribute only to a part of the spectrum, starting from the minimum energy  up to around 16 MeV, which corresponds to the $\alpha$ particle reaching the daughter nucleus potential barrier at exactly the height of the initial energy $E_0$. The recollision contribution  can be evaluated by comparing the spectra with and without recollisions. In order to qualitatively describe the case of no recollisions for the same laser field strength, we have considered the case when the Coulomb barrier is removed once the particle has tunneled through. Thus, the dynamics of the $\alpha$ particle outside the barrier is dictated solely by the laser field and no recollision is possible.
Comparing qualitatively the two spectra,  the recollision plateau is clearly visible, having an extension which corresponds to approx.   6$U_p$. 
The sharp peaks occur where two trajectories meet at the classical cut-offs and produce a divergency. This is a well-known issue in strong-field atomic physics  and has been widely investigated \cite{Paulus,Carla,Sergey}.
In our calculation this divergency is fixed by taking into account the third derivative of the action in the expression of the saddle points and determining the regularization factor $\beta$ in Eq.~(\ref{transitionamplitudemonochromatic_alpha}) as described in Ref.~\cite{Sergey}. The peaks occur at the spectrum boundaries, meeting point for the  two regular trajectories,  at around 2  MeV, where a regular and a recollision
trajectory meet and at 6 MeV where the two recollision trajectories meet.  The rising  shoulder up to approx. 13 MeV appears as one of the two  recollision trajectories becomes the dominant contribution of the imaginary part of the classical action. The recollision plateau in the energy spectrum of the detected $\alpha$ particles with its characteristic features can serve as a clear signature for the occurrence of recollisions.

A list of suitable $\alpha$ emitters for laser-driven recollisions can be found in Ref.~\cite{Buck}. The half-lives of the $\alpha$ emitters span between $10^{-7}$ and $10^{17}$ s. Short-lived parent nuclei have the advantage that a measurable fraction of them will decay during the laser pulse duration of 500 fs. With an appropriately large number of parent nuclei originally present, one can reach $\alpha$ decay rates per pulse of 10--100. For this we need to bring $10^7-10^8$ parent nuclei with a half-life  of $10^{-7}$ s such as $^{212}\mathrm{Po}$ into the laser focus of $10^{-3}$ cm$^2$. Such short-lived nuclei are usually produced in nuclear reactions and then separated in-flight as for instance at the Fragment Separator at the GSI facility in Darmstadt, Germany \cite{FRS,SIS100}. An additional possible signature of laser-driven recollisions can be seen in the bremsstrahlung spectrum emitted during $\alpha$ decay. For long-lived parent nuclei, solid-state samples with high density theoretically ensure a large number of $\alpha$ emitters in the laser focus. The disadvantage of such a setup is that solid-state targets can only be manufactured of isotopes with a long half-life,  $t_{1/2}\gg 10^{7}$ s, leading to  a rate of only approx. 0.1  decays per pulse even for as much as $10^{19}$ nuclei located in the laser focus. Furthermore, the impact of a laser beam with intensities of $10^{22}-10^{23}$ W/cm$^2$ on an overdense solid-state target will lead to strong perturbing effects such as screening,  Coulomb interaction with neighboring nuclei and sample destruction  on time scales faster than the nuclear half-lives. These effects drastically reduce the recollision  probability.

With increasing field intensity, the ponderomotive energy can reach values which allow the recolliding $\alpha$ particle to penetrate the nuclear barrier. 
For this typically energies of about 20 MeV or more are required. 
A study of the nuclear reaction databases \cite{exfor} reveals that a number of daughter nuclei of $\alpha$ emitters are known to undergo nuclear reactions when bombarded by energetic $\alpha$ particles. 
Typical possible nuclear reactions are inelastic scattering $(\alpha,\mathrm{inl})$ (for 24 MeV $\alpha$ projectiles interacting with $^{182}\mathrm{W}$ or 13--24 MeV $\alpha$ projectiles interacting with $^{184}\mathrm{Os}$ or $^{182}\mathrm{Os}$, for instance), neutron emission $(\alpha,2\mathrm{n})$ at 20 MeV  for $^{208}\mathrm{Pb}$ or even fission $(\alpha,\mathrm{f})$ for 20 MeV $\alpha$ particles rescattered on $^{226}\mathrm{Ra}$. 
All the listed isotopes are reached via $\alpha$ decay of parent nuclei. 
While such nuclear reactions are already available experimentally, the novelty in laser-induced recollisions in $\alpha$ decay manifests in the time scale of the process. The recolliding particles are emitted during a specific phase interval of each laser cycle only. Thus,  recollisions occur on a 1 fs interval, allowing the alpha particle to probe the daughter nucleus on a much shorter time scale than  available in experiments using ion beams.
Of particular interest are the cases when the daughter nucleus is partially produced in an excited state such as the $\alpha$ decays of $^{212}\mathrm{Po}$ or $^{241}\mathrm{Am}$.  The recolliding $\alpha$ particles then probe on a fs scale  the 2.6 MeV  excited state of $^{208}\mathrm{Pb}$  ($t_{1/2}^0=16.7$ ps) or the low-lying 59.5 keV ($t_{1/2}^0=67$ ns) and 102.9 keV ($t_{1/2}^0=80$ ps) states of $^{237}\mathrm{Np}$. Laser-driven recollisions could in these cases coherently trigger a variety of nuclear reactions and probe the relaxation dynamics of the daughter nucleus after $\alpha$ decay on an unprecedented femtosecond scale. 

Due to the similarities in the theoretical description, one should keep in mind that laser-driven recollisions may occur also in the case of laser-assisted proton emission. 
Proton emitters are usually nuclei very far from the line of stability on the proton-rich side \cite{Blank2008,Pfu2012}, thus offering the possibility to investigate the complementary region of the neutron-rich $\alpha$-emitters on the nuclear chart.  
The  half-lives of proton emitters can be comparable or shorter than the ones for $\alpha$ decay. From the theory side,  the
calculation of proton-decay rates is usually performed via the straightforward application of the $\alpha$-decay theory albeit with the simplification that there is no need to consider the preformation factor \cite{Buck,Buck92}. Just like the $\alpha$ particle, the proton tunnels the spherical Coulomb and the centrifugal barrier created by its interaction with the core nucleons of the  daughter nucleus. As a new feature, the role of the centrifugal barrier in
proton emission is more important than in $\alpha$ decay due to the smaller proton reduced mass and also because in
most cases the proton is originally unpaired in the parent nucleus and  carries a non-vanishing angular momentum.
Due to a different charge/mass ratio for proton and $\alpha$ particle, the dynamics of the tunneled particle outside the barrier and the recollision spectra will be quantitatively different (although, as an interesting feature, the ponderomotive energy is the same for both proton and $\alpha$ particle), but the main recollision features should be reproduced. Due to the high degree of exoticism of proton-emitting nuclei, little is known about the nuclear reactions cross-sections that can occur for protons rescattering on the daughter nucleus. A study of laser-driven proton recollisions might therefore provide important information for this nuclear parameter region so far from stability.

The authors would like to thank Karen Z. Hatsagortsyan for fruitful discussions. 


\end{document}